\documentclass[12pt]{iopart}
\usepackage{iopams}
\usepackage{setstack}
\usepackage{graphicx,color}
\usepackage{ulem}
\begin{document}

\title{Optimal linear Kawasaki model}

\author{Shaon Sahoo}
\address{Solid State and Structural Chemistry Unit, Indian
Institute of Science, Bangalore 560012, India}

\author{Sandeep Chatterjee}
\address{School of Physical Sciences, National Institute of Science Education 
and Research, Bhubaneshwar 751005, India}

\begin{abstract}
The Kawasaki model is not exactly solvable as any choice of the 
exchange rate ($w_{jj'}$) which satisfies the detailed balance 
condition is highly nonlinear. In this work we address the issue of 
writing $w_{jj'}$ in a best possible linear form such that the mean 
squared error in satisfying the detailed balance condition is least. 
In the continuum limit, our approach leads to a Cahn-Hilliard equation 
of conservative dynamics. The work presented in this paper will help us 
anticipate how the conservative dynamics of an arbitrary Ising 
system depends on the temperature and
the coupling constants. In particular, for two and three dimensional 
systems, the critical temperatures estimated in our work are in good 
agreement with the actual values. We also calculate the dynamic and 
some of the critical exponents of the model.
\end{abstract}

%\maketitle

\section{Introduction}
From an arbitrary given state how does a system relax to its equilibrium 
state? Answering this question remains one of the major challenges in 
theoretical physics. Even after a century of rigorous research, our 
understanding of nonequilibrium systems is modest. 
Though we have a well defined prescription to address 
any issue regarding an equilibrium state, we are yet to develop a general 
framework to study some irreversible process. In this challenging situation, 
it is important to study the simple physical models in order to gain some 
insight into the nonequilibrium processes.

The Ising model is one of the simplest non-trivial models to study the
underlying physics of many irreversible processes. Here the main two 
microscopic mechanisms, by which a system equilibrates, are due to 
Roy J. Glauber \cite{glauber63} and K. Kawasaki \cite{kawasaki66}. While 
the first mechanism (Glauber's) describes a non-conservative order-parameter 
dynamics, the second one is thought to be the underlying mechanism for a 
conservative order-parameter dynamics. The Glauber dynamics is exactly 
solvable only for one dimensional Ising system, on the other hand, the 
Kawasaki dynamics is not exactly solvable even for one dimensional system. 
Though the choice of transition rate or exchange rate for the two kinetic 
models are not unique, it has to be such that the detailed balance condition 
at equilibrium is satisfied. Although it is possible 
to get some nonlinear form of the transition rate or exchange rate for which 
the detailed balance condition is satisfied, unfortunately, the kinetic 
models are not exactly solvable with this nonlinear form (except for the 
aforementioned case). 

It has been a real challenge to study analytically these two kinetic models 
for an arbitrary Ising system (in any dimension) without compromising on the 
detailed balance condition. Recently one of us (with another author) developed 
a general mathematical method to study the Glauber dynamics in an arbitrary 
Ising system \cite{sahoo14}. It may be noted that, the linear Glauber model, 
where the chosen transition rate is linear, is exactly solvable although the 
detailed balance condition is not exactly satisfied 
\cite{scheucher88,oliveira03,hase06}. In our mathematical approach 
\cite{sahoo14}, a 
linear form of the transition rate with an appropriate number of parameters 
is taken. These parameters are then optimized in such a way that the mean 
squared error in satisfying the detailed balance condition is least. The 
advantage of this method is that, it helps us to anticipate how the kinetic 
properties of an Ising system depend on the temperature and coupling constant.  
Along with many other things, it was shown in that work that, using the method 
it is possible to derive a time-dependent Ginzburg-Landau equation (linear 
version) for the non-conservative dynamics from the Glauber's microscopic 
model \cite{sahoo14}. In this paper, we use this optimal linearization approach 
to study the Kawasaki dynamics in an arbitrary Ising model. It will be shown here 
how a Cahn-Hilliard equation (linear version) for the conservative dynamics can 
be derived from the Kawasaki's microscopic model. It is
very encouraging to find 
that the critical temperatures (for two and three dimensional systems) 
estimated from the divergence of the correlation length or the critical slow 
down are in very good agreement with the 
actual values. We also calculate the dynamic exponent and some of the 
critical exponents for our optimal linear model.

Our paper is organized in the following way. In section 2, we give a
detailed description of our approach. In the next section (sec 3),
we apply our method to study Ising systems in different dimensions. 
We conclude our work in section 4.
 
\section{General theory}
Let us consider an Ising system of $N$ interacting spins 
($\sigma_i = \pm 1$). For simplicity, we will consider in this work 
a uniform (or isotropic) ferromagnetic system where all the coupling 
constants are same. It will be also assumed that there is only nearest 
neighbor interactions and the system is on a hyper-cubic lattice. 
The relevant Hamiltonian is given by,
\begin{eqnarray}
H = -J\sum_{<jj'>}\sigma_j\sigma_{j'}, 
\label{isngham}
\end{eqnarray}
where $J$ is the coupling constant ($> 0$). We further consider that 
$p(\{\sigma\}; t)$ is the probability that
the spins take the values $\sigma_1, \cdots, \sigma_N$ at time $t$.
A master equation for the time evolution of the probability is given 
by, 
\begin{eqnarray}
\frac{d}{dt}p(\{\sigma\}; t) = -\sum_{<jj'>} w_{jj'}(\sigma_j\sigma_{j'}) 
p(\{\sigma\}; t)  
+ \sum_{<jj'>} w_{jj'}(\sigma_{j'}\sigma_j) p(\{\sigma\}_{jj'}; t),
\label{mstreq} 
\end{eqnarray}
where summations run over all possible nearest neighbor pairs. 
$\{\sigma\}_{jj'}$ represents the same state as $\{\sigma\}$ with spins
of the pair $jj'$ exchanged (i.e., $\sigma_j \rightarrow  \sigma_{j'}$ and 
$\sigma_{j'} \rightarrow \sigma_j$). In 
Kawasaki dynamics, the exchange rate $w_{jj'}(\sigma_j\sigma_{j'})$ 
for the neighboring 
pair $jj'$ is the transition rate from the state $\{\sigma\}$ to the 
state $\{\sigma\}_{jj'}$. 

By considering $\sigma_j(t)$ as stochastic function of time, we  
consider two important quantities, namely, a time dependent 
average spin value $q_j(t)$ and a time dependent correlation function 
$r_{i,j}(t)$. These are given below,
\begin{eqnarray}
\label{avq}
q_j(t)& =& \langle \sigma_j(t) \rangle = \sum_{C(N)} \sigma_j 
p(\{\sigma\}; t) \\
\label{avr}
r_{i,j}(t)& =& \langle \sigma_i(t)\sigma_j(t) \rangle = \sum_{C(N)} 
\sigma_i \sigma_j p(\{\sigma\}; t).
\end{eqnarray}
Here summation is over all possible ($2^N$ in number) spin configurations,
$C(N)$. It may be noted that $r_{j,j} = 1$.

We now write time derivative of these quantities as first step to obtain
them as function of time. It is easy to get the derivatives of $q_j$ and 
$r_{i,j}$ by multiplying respectively $\sigma_k$ and $\sigma_j\sigma_k$ 
to Eq. (\ref{mstreq}) and then sum
them over all possible spin configurations. Some easy manipulations would 
give us the following equations:
\begin{eqnarray}
\label{avq_t}
\frac{d}{dt}q_k(t) =-2\sum_{C(N)} \sum_{k'=L_k}\sigma_k w_{kk'}
(\sigma_k\sigma_{k'}) p(\{\sigma\};t)\\
\label{avr_t}
\frac{d}{dt} r_{j,k} (t) = - 2\sum_{C(N)} \sigma_j \sigma_k 
\left\{\sum_{j'=L_j}w_{jj'}(\sigma_j\sigma_{j'})\right. \nonumber \\
~~~~~~~~~~~~~~~~~~~~~~~~~~~~~~~~~~~ 
\left. + \sum_{k'=L_k} w_{kk'}(\sigma_k\sigma_{k'})\right\} 
p(\{\sigma\};t).
\end{eqnarray}
Here, for example, the second summation in Eq. (\ref{avq_t}) is over 
all neighbors of the $k$th site (denoted by $L_k$).

To solve these equations, a choice of $w_{jj'}$ has to be made.
The exchange rate 
$w_{jj'}$ should be chosen in such a way that it satisfies the equation 
of detailed balance (EDB) at the equilibrium. In addition, it should 
be zero when neighboring pairs $jj'$ are both up or both down. 
We here use Suzuki-Kubo
form for $w_{jj'}$ which satisfies the EDB at equilibrium \cite{suzuki68,puri09}. 
To make sure that the rate is zero when both the spins are aligned along 
the same direction, we multiply $w_{jj'}$ by a factor
$\frac 12 (1-\sigma_j\sigma_{j'})$. So our chosen form for $w_{jj'}$ is 
given by,
\begin{eqnarray}
w_{jj'}(\sigma_j\sigma_{j'})=\frac{\alpha}{2}\left[1-{\rm tanh}
\left(\frac{\beta \Delta E}{2}\right)\right]\cdot\frac 12 
(1-\sigma_j\sigma_{j'}),
\label{exchrate} 
\end{eqnarray}
with $\Delta E$ being the energy difference between the final state 
$\{\sigma\}_{jj'}$ and the initial state $\{\sigma\}$. Here $\beta$ is 
the inverse temperature $1/(k_BT)$ with $k_B$ being the Boltzmann 
constant. The parameter $\alpha$ 
sets the timescale of the nonequilibrium process. It is not difficult 
to see that, 
\begin{eqnarray}
\Delta E = J(\sigma_j - \sigma_{j'})\left(\sideset{}{'}\sum_{m = L_j}
\sigma_m-\sideset{}{'}\sum_{n = L_{j'}}\sigma_n\right).
\label{ergdff}
\end{eqnarray}
Here the first (second) primed summation runs over all the neighbors of 
$j$th ($j'$th) site excluding the $j'$th ($j$th) site. Now using the 
above expression of $\Delta E$ in Eq. (\ref{exchrate}), we get,
\begin{eqnarray}
\label{exchrate1} 
w_{jj'}(\sigma_j\sigma_{j'})=\frac{\alpha}{2}\left[
\vphantom{\sideset{}{'}\sum_{n = L_{j'}}\sigma_n)}\frac 12 
(1-\sigma_j\sigma_{j'})\right. \nonumber \\
~~~~~~~~~~~~~~~~~~~~~~\left. -\frac 12 (\sigma_j - \sigma_{j'}){\rm tanh}
\left\{\beta J \left(\sideset{}{'}\sum_{m = L_j}
\sigma_m-\sideset{}{'}\sum_{n = L_{j'}}\sigma_n\right)\right\}\right].
\end{eqnarray}

Unfortunately, this exact nonlinear form of $w_{jj'}$ is intractable 
for the analytical 
study of dynamics. A linear form of $w_{jj'}$ is easy to handle, but, 
generally it does not exactly satisfy the detailed balance condition. 
We now present a mathematical approach to linearize $w_{jj'}$ in such 
a way that the mean squared error in satisfying the detailed balance 
condition is least.

\subsection{Linearization of $w_{jj'}$ using a least squares method}
In this subsection we will see how one can linearize the Suzuki-Kubo 
form of the exchange rate in an optimal way. More specifically, we 
will discuss here the best possible way to linearize 
hyperbolic-tan function appearing in $w_{jj'}$; this will in turn  
ensure that, the error in satisfying the EDB is least \cite{sahoo14}.
The remaining nonlinearity in $w_{jj'}$ due to 
the constraint term $\frac 12 (1-\sigma_j\sigma_{j'})$ will be 
systematically handled in section 3. 

Let us consider a hyperbolic-tan function of Ising variables 
$tanh~[\sum_k p_k \sigma_k]$, where $p_k$ are some dimensionless real 
parameters. Noting the series ${tanh}~x= x -\frac{x^3}{3} + \cdots$, 
we can attempt to linearize our hyperbolic-tan function by considering,
\begin{eqnarray}
\label{mpexpr}
{\rm tanh}\left[\sum_{k=1}^{L} p_k \sigma_k\right] \approx 
\sum_{k=1}^L\gamma_k \sigma_k.
\end{eqnarray}
Here the coefficients $\gamma_k$'s are not just $p_k$'s that appear in the
first order term of the hyperbolic-tan series. These coefficients also have
contributions
from the higher order terms of the series (this will be clear by noting that,
$\sigma_j^m=1$ if $m$ is even and $\sigma_j^m=\sigma_j$ if $m$ is odd). Although
by analyzing the series it is possible to find out the exact values of $\gamma_k$'s,
it is best to take the optimal values for the $\gamma_k$'s which can be obtained
by a linear regression process.
By taking the optimal values, we ensure that the error introduced due to 
linearization is minimum. The
optimization process somewhat compensates the absence of the nonlinear terms in
our desired linear form of $w_j$ (nonlinear terms are typically product of
different $\sigma$'s).

To do a linear regression, we will consider $\gamma_k$'s in Eq. (\ref{mpexpr}) as
the parameters of the regression process. We may note that, Eq. (\ref{mpexpr})
actually represents $2^L$ linear equations in $L$ parameters. Each of these linear
equations corresponds to the one of the $2^L$ configurations of the $L$ Ising 
variables. 
Obviously, no set of values for the $\gamma$'s can simultaneously satisfy the
overdetermined set of $2^L$ linear equations.
We will now see how the best possible values for $\gamma$'s, for which mean 
squared error is minimum, can be obtained.

Before discussing the linear regression process, it may be worth mentioning here
that, the function $tanh~x$ is linear about the origin ($x = 0$). Since the term
$[\sum_{k=1}^{L} p_k \sigma_k]$ is zero or close to zero for a good fraction
of the total number of configurations (at least for isotropic case when $p_k$'s are
equal), we expect our linearization to work reasonably good in a normal situation.
%Other way to argue in favor of this linearization is that, during coarsening,
%the Kawasaki dynamics happens only at the domain boundaries (not inside a domain) 
%where $\sum_{k=1}^{L} \sigma_k$ is generally zero or close to zero. Inside a domain 
%this linearization will not work good, but we do not have to bother about that as 
%the exchange rate $w_{jj'}$ is anyway zero inside a domain 
%(see Eq. (\ref{exchrate1})).
 
When all $p_k$'s are different, we need to consider $L$ number of independent 
parameters ($\gamma$'s). A Moore-Penrose pseudoinverse 
matrix of dimension $L\times 2^L$ can be used to get the best possible values of 
the parameters ($\gamma$'s). This pseudoinverse matrix involved in the regression 
process is obtained solely from the configuration matrix (whose different rows 
represent different configurations of the $L$ Ising spins) and does not depend on 
any parameter of the problem. A general discussion on this topic can be found 
in Ref. \cite{sahoo14}. We will consider a special case here. 
Since our system is isotropic (all the coupling constants are same) with only 
nearest neighbor interactions, values of all the parameters obtained 
in the regression process will be effectively same. 
Therefore consideration of a single parameter in the regression process is 
good enough 
for the present purpose; let this parameter be $\gamma$. 
The regression process is now reduced to finding 
the best possible value of the parameter from the following set of equations:
\begin{eqnarray}
\label{lrkwski} 
{\rm tanh}\left\{\beta J \left(\sideset{}{'}\sum_{m = L_j}
\sigma_m-\sideset{}{'}\sum_{n = L_{j'}}\sigma_n\right)\right\} 
\approx \gamma \left( \sideset{}{'}\sum_{m = L_j}
\sigma_m-\sideset{}{'}\sum_{n = L_{j'}}\sigma_n\right).
\end{eqnarray}
If $L$ be the number of Ising variables involved in the above expression, 
we may note that the above expression actually represents $2^L$ equations 
corresponding to each of that many configurations. Here, if $z$ is the 
coordination number or the number of nearest neighbors 
(2, 4, and 6 respectively for one, two and three dimensional systems), then 
$L = 2(z-1)$. In the regression process, the left side of Eq. (\ref{lrkwski}) 
will be represented by a column matrix with $2^L$ elements; let us denote this 
column by $\Omega$. Similarly, the quantity inside the bracket in the right side 
of Eq. (\ref{lrkwski}) will again be represented 
by a  column matrix with $2^L$ elements; we will denote this column by $C$ matrix. 
An error function can now be defined from these two column matrices: $S(\gamma) = 
\sum_{i=1}^{2^L}|\Omega_i-\gamma C_i|^2= ||\Omega-\gamma C||^2$. The best possible 
value of the parameter $\gamma$ can be obtained by minimizing the error function 
$S(\gamma)$. The formal solution for $\gamma$ can be written using the 
Moore-Penrose pseudoinverse matrix $C^+ = (C^TC)^{-1}C^T$, which in the present 
case is just a row matrix with $2^L$ number of elements. The solution is given by 
the following relation:
\begin{eqnarray}
\label{gamma_sol} 
\gamma = C^+\Omega.
\end{eqnarray}

To get the exact expression of $\gamma$ in terms of the parameters $\beta$ and 
$J$, we note that, among the $2^L$ number of elements of the column matrix $C$, 
one element will be $L$ and, due to $Z_2$ symmetry, $-L$ will be another element. 
There will be $^LC_1$ number of elements with the value $L-2$ and equal number 
of elements with the value $-L+2$. 
This counting goes on till we get $^LC_{\frac L2}$ number of 0's (note, $L$ is 
always even). For us $C^TC$ is just a number whose value is $L2^L$. The 
pseudoinverse matrix  in the present case is given by, 
$C^+ = \frac{2^{-L}}{L}C^T$. We notice that if $i$th element of the $C$ matrix 
is, say, $Y$, then the $i$th element of the $\Omega$ matrix will be 
$tanh~(\beta JY)$. Now using Eq. (\ref{gamma_sol}), it is easy to get the 
desired expression for the $\gamma$:
\begin{eqnarray}
\label{gamma_value} 
\gamma = \frac{2^{-L+1}}{L}\sum_{i=1}^{L/2}Z_i~ ^LC_{i-1} {\rm tanh}
[\beta J Z_i],
\end{eqnarray}
where $Z_i = L-2i+2$ (here we again remember that $L = 2(z-1)$).

Using this optimal linearization of hyperbolic-tan function, we can rewrite 
the exchange rate give in Eq. (\ref{exchrate1}) as,
\begin{eqnarray}
\label{exchrate2} 
w_{jj'}(\sigma_j\sigma_{j'})=\frac{\alpha}{4}\left[
\vphantom{\sideset{}{'}\sum_{n = L_{j'}}\sigma_n)} 
(1-\sigma_j\sigma_{j'}) %~~~~~~~~~~~~~~~~~~~~~~~~\\ \nonumber 
-\gamma (\sigma_j - \sigma_{j'})
\left(\sideset{}{'}\sum_{m = L_j}
\sigma_m-\sideset{}{'}\sum_{n = L_{j'}}\sigma_n\right)\right].
\end{eqnarray}

It may be worth commenting here about the nature of the steady state that 
one would get by using the above exhange rate. We may note that, to satisfy 
the detailed balance condition, the local probabilty current for any pair $jj'$, 
$I_{jj'} = - w_{jj'}(\sigma_j\sigma_{j'}) 
p(\{\sigma\}) +  w_{jj'}(\sigma_{j'}\sigma_j) p(\{\sigma\}_{jj'})$, 
should be zero for every configuration 
of its neighbours (here $p(\{\sigma\})$ is the Maxwell-Boltzmann 
probability factor defined for the configuration $\{\sigma\}$). Had we taken 
the nonlinear form for $w_{jj'}$, as given in Eq. (\ref{exchrate1}), the current 
$I_{jj'}$ would have been zero for the every configuration
of its neighbours. With the exchange rate given in Eq. (\ref{exchrate2}), the 
current will not be zero for every configurations -sometimes it will be positive 
and sometimes negative. In this context, as explained in Ref. \cite{sahoo14}, 
our method ensures following things: (a) the average local probability current 
$<I_{jj'}>$ (average over all possible configurations of neighbors) is zero, and 
(b) two opposite tendencies (forward currents and backward currents depending on 
the sign of $I_{jj'}$) are individually as low as possible on the average.

\section{Study of dynamics in continuum limit}

A continuum approach will be adopted in this section to study the Kawasaki 
dynamics. We will see that in this limit both the equation for the average 
local spin and the equation for the correlation function (see 
Eqs. (\ref{avq_t}) and (\ref{avr_t})) take the same form with parameters of 
the equations differ only by a factor of 2.

\subsection{Equation for local magnetization}
We will first consider the one dimensional system. Using the exchange rate 
from Eq. (\ref{exchrate2}), we get from Eq. (\ref{avq_t}):
\begin{eqnarray}
\label{Q_eq} 
\frac{d}{dt}q_k(t) = -\frac{\alpha}{2}\left[\left(2q_k-q_{k-1}-q_{k+1}
\right)\right.\nonumber\\
-\gamma \left(q_{k+1}+q_{k-1}-q_{k+2}-q_{k-2}\right) \nonumber\\
\left. -\gamma \left(-2 <\sigma_{k-1}\sigma_k\sigma_{k+1}> + 
<\sigma_{k-2}\sigma_{k-1}\sigma_{k}>
+<\sigma_{k}\sigma_{k+1}\sigma_{k+2}>\right)\right].
\end{eqnarray}

Now in the continuum limit if $Q(x,t)$ denotes the local magnetization at the 
location $x$ and time $t$, then the first group of terms in the right hand side, 
i.e. $2q_k-q_{k-1}-q_{k+1}$, can be recognized as $-\frac{d^2}{dx^2}Q(x,t)$. 
Similarly, the second group of terms, i.e. $q_{k+1}+q_{k-1}-q_{k+2}-q_{k-2}$, 
can be recognized as $-3\frac{d^2}{dx^2}Q(x,t)$. Last or third group of terms, 
where all the terms are three-point correlation functions, is difficult to deal 
with. 
%If we write $S(x,t) = <\sigma_{k-1}\sigma_k\sigma_{k+1}>$, then the third 
%group of terms 
%can be recognized as $\frac{d^2}{dx^2}S(x,t)$. 
To make the calculations tractable, we will replace the nonlinear terms like 
$\sigma_{k-1}\sigma_k\sigma_{k+1}$ by a suitable linear form. Let us consider 
that the Ising system is momentarily fixed (say, at time $t$). This `frozen' 
system will have domains of `up' and `down' spins. We now note here that, 
when all the three spins, i.e. 
($k-1$)th, $k$th and ($k+1$)th spins, are from the same domain, they will be 
aligned along the same direction. The value of the product of these spins will 
be same as the value of a single spin. That is to say, 
$\sigma_{k-1}\sigma_k\sigma_{k+1} = \sigma_k$ if all the three spins are from 
the same domain. 
If ($k-1$)th and $k$th spins are from one domain and ($k+1$)th 
spin is from the next domain, then the above linearization will not work. 
Similarly, this linearization also breaks down when ($k-1$)th spin is in one 
domain and the other two spins are from the next domain. This indicates that we 
need to consider one more term whose value is zero inside a domain and which 
appropriately adjusts the boundary effects. A careful inspection shows that the 
term ($\sigma_{k-1} - 2\sigma_k + \sigma_{k+1}$) fulfils our requirement. 
So we replace $\sigma_{k-1}\sigma_k\sigma_{k+1}$ by the following linear 
term: $\sigma_k + (\sigma_{k-1} - 2\sigma_k + \sigma_{k+1})$. 
This linearization will not only be 
valid inside a domain but also at boundaries. More specifically, above 
linearization is exact for the following four configurations: (1,1,1), 
(-1,-1,-1), (1,1,-1) and (-1,-1,1). Here it should be also mentioned that this 
linearization does not work for an isolated spin in a domain (like an up spin 
inside a down spin domain), i.e., it fails for the remaining two configurations
(1,-1,1) and (-1,1,-1). However it can be  
safely assumed that, after some time in 
the coarsening, the number of such isolated spins becomes extremely small 
compared to the total number of spins $N$ (which is assumed to be 
thermodynamically large). Here we may  
recall that in a spin exchange 
dynamics there is zero chance that a spin inside a domain will suddenly flip to 
become such an isolated spin. 
Above argument shows that the error introduced due to the linearization 
of the three-point correlation function can be assumed to be very small. 
In the continuum limit, if we denote  
$<\sigma_{k-1}\sigma_k\sigma_{k+1}>$ by $S(x,t)$, then the linearization allows 
us to write $S(x,t) = Q(x,t) + \frac{d^2}{dx^2}Q(x,t)$. The third group of
terms in the right hand side of Eq. (\ref{Q_eq}) can now be written as 
$\frac{d^2}{dx^2}S(x,t)$ or, $\frac{d^2}{dx^2}Q(x,t) 
+ \frac{d^4}{dx^4}Q(x,t)$.

With the continuum limit representation of all the terms in Eq. (\ref{Q_eq}), 
we now write the equation for the local magnetization in the continuum limit:
\begin{eqnarray}
\label{Q_eq1} 
\frac{\partial}{\partial t} Q(x,t)=\frac{\alpha}{2}(1-2\gamma)
\frac{d^2}{dx^2} Q(x,t) -\frac{\alpha}{2}\gamma\frac{d^4}{dx^4} Q(x,t).
\end{eqnarray}

Using the same line of arguments, it is possible to generalize this equation to 
the higher dimensions. In dimension $d$, the equation reads in the following way,
\begin{eqnarray}
\label{Q_eq_d} 
\frac{\partial}{\partial t} Q(\vec{r},t)=D \nabla^2 Q(\vec{r},t) - \kappa  
\nabla^4 Q(\vec{r},t).
\end{eqnarray}
Here $Q(\vec{r},t)$ is the local magnetization at the location $\vec{r}$ and time $t$. 
In the above equation, the diffusion constant $D = \frac{\alpha}{2}(1-2d\gamma)$ and  
the bidiffusion constant $\kappa = \frac{\alpha}{2}\gamma$. 
The term $d$ represents the dimensionality of the system (i.e., $d$ = 1, 2 or 3 for 
one, two or three dimensional system respectively). The parameter 
$\gamma$ depends on dimensionality $d$ (=$z/2$) and is given by Eq. (\ref{gamma_value}).
%\textcolor{red}{(Here can we comment on how this compares with the continuum version of
%non conservative dynamics...the new term..its possible effect..)}

After adopting all these linearization approximations, one would like to know if the 
dynamics still remains conservative. It is in fact very easy to check from 
Eq. (\ref{Q_eq}). After 
replacing three point correlation terms by their appropriate linear versions, if we 
sum the terms of Eq. (\ref{Q_eq}) over all the sites, we will get 
$\frac{dM(t)}{dt} = 0$, where $M(t)$ is the total magnetization 
(= $\sum_{k=1}^N q_k$). This shows that $M(t)$ remains constant over time as expected 
for a conservative dynamics. This fact can also be explicitly checked for higher 
dimensional systems. 

Eq. (\ref{Q_eq_d}) is in the form of the well known Cahn-Hilliard equation 
(linear version) of the conservative dynamics \cite{cahn59,cahn65}, which has 
been subject of 
active study for the last few decades \cite{krapivsky10,bray94,puri09}. The work 
presented in this paper thus establishes a connection between the phenomenologiacl 
Cahn-Hilliard theory and the Kawasaki's microscopic model for conservative dynamics. 
Advantage of the present work is obvious; it gives us explicit temperature and 
exchange constant dependence of the parameters involved in the Cahn-Hilliard equation. 
For two and three dimensional systems, the critical temperatures estimated by 
analyzing this equation are in very good agreement with the well know actual 
values (section 3.3). Here it may be further added that, if the equation for 
a non-conservative order parameter $Q$ is of the form $\frac{\partial}{\partial t} Q 
= F(Q)$, then the corresponding equation for the case where $Q$ is a conserved order 
parameter is given by: $\frac{\partial}{\partial t} Q = - 
\nabla^2 F(Q)$ \cite{krapivsky10}. In this respect, 
Eq. (\ref{Q_eq_d}) is consistent with the time-dependent Ginzburg-Landau equation 
derived from the Glauber's model in our previous work \cite{sahoo14} 
within the same optimal linearization approximation.

\subsection{Equation for two-point correlation}
Similar to the treatment of the local magnetization, we will first consider here 
the one dimensional system. Using the exchange rate from Eq. (\ref{exchrate2}), 
we get from Eq. (\ref{avr_t}):
\begin{eqnarray}
\label{G_eq} 
\frac{d}{dt}r_{j,i}(t) = -\alpha\left[\left(2r_{j,i}-r_{j,i-1}-r_{j,i+1}
\right)\right. \nonumber \\
-\gamma \left(r_{j,i+1}+r_{j,i-1}-r_{j,i+2}-r_{j,i-2}\right) \nonumber \\
\left. -\gamma \left(-2 <\sigma_j\sigma_{i-1}\sigma_i\sigma_{i+1}> + 
<\sigma_j\sigma_{i-2}\sigma_{i-1}\sigma_{i}>
+<\sigma_j\sigma_{i}\sigma_{i+1}\sigma_{i+2}>\right)\right]
\end{eqnarray}

Now in the continuum limit if $G(x,t)$ denotes the time dependent correlation 
between two spins 
separated by a distance $x$ ($x = |i-j|$), then the first group of terms, i.e. 
$2r_{j,i}-r_{j,i-1}-r_{j,i+1}$, can be recognized as 
$-\frac{d^2}{dx^2}G(x,t)$. Similarly, the second group of terms can be recognized 
as $-3\frac{d^2}{dx^2}G(x,t)$. Last or third group of terms, where all
the terms are four point correlation functions, is again difficult to deal with. 
If we now adopt the linearization discussed in the last subsection, i.e., if we 
replace the terms like $\sigma_{i-1}\sigma_i\sigma_{i+1}$ by the terms like 
$\sigma_i + (\sigma_{i-1} - 2\sigma_i +\sigma_{i+1})$, then the third term in 
Eq. (\ref{G_eq}) can be recognized as $\frac{d^2}{dx^2}G(x,t) 
+ \frac{d^4}{dx^4}G(x,t)$. 
This enables us to write the equation for the 
correlation function in the following way:
\begin{eqnarray}
\label{G_eq1} 
\frac{\partial}{\partial t} G(x,t)=\alpha(1-2\gamma) 
\frac{d^2}{dx^2} G(x,t) - \alpha \gamma \frac{d^4}{dx^4} G(x,t).
\end{eqnarray}

It is again straightforward to generalize this equation for the higher 
dimensional systems. The result is given below,
\begin{eqnarray}
\label{G_eq_d} 
\frac{\partial}{\partial t} G(r,t) = 2D \nabla^2 G(r,t) - 
2\kappa \nabla^4 G(r,t).
\end{eqnarray}
Here $G(r,t)$ is the correlation function between the sites separated by a 
distance $r$ (= $|\vec{r}|$). 
In the equation, we have again $D = \frac{\alpha}{2}(1-2d\gamma)$ and 
$\kappa = \frac{\alpha}{2}\gamma$.
Physically appealing general solution of this equation is difficult. 
We will only consider its steady state solution in section 3.4.

\subsection{Dynamical exponent, correlation length and critical temperature} 
To gain some insight into Eq. (\ref{Q_eq_d}), we will do a Fourier analysis of 
the equation. This analysis will give us the equation for each mode $\vec{k}$:
\begin{eqnarray}
\label{fourier} 
\frac{\partial Q(\vec{k},t)}{\partial t} = -k^2(D + \kappa k^2)Q(\vec{k},t),
\end{eqnarray}
%where, $D = \frac{\alpha}{2}(1 - 2d \gamma)$ and 
%$\kappa = \frac{\alpha}{2} \gamma$. 
Solution of this equation gives, 
$Q(\vec{k},t) = Q(\vec{k},0){\rm e}^{-t/\tau (\vec{k})}$, where the 
relaxation time for the mode $\vec{k}$ is given by:
\begin{eqnarray}
\label{rlxn} 
\tau (\vec{k})^{-1} = \kappa k^2(k^2 + \xi^{-2}).
\end{eqnarray}
Here the correlation length $\xi = \sqrt{\frac{\kappa}{D}} = 
\sqrt{\frac{\gamma}{1 - 2d \gamma}}$. 
%\textcolor{red}{(Then we have the 
%complete solution, may not be in a helpful state)}

The dynamic exponent (denoted by z; not to be confused with coordination
number) is defined by how the maximum possible value of the relaxation time 
($\tau_{\rm max}$) scales with the system's relevant length scale. If we 
consider a finite but large system of size $L$, then $k_{\rm min} \sim 1/L$. 
Now when $k \ll \xi^{-1}$, then $\tau_{\rm max} \sim k_{\rm min}^{-2}\xi^2$, 
i.e., $\tau_{\rm max} \sim L ^2 \xi^2$. In this limit, both the length scales 
($L$ and $\xi$) are relevant; individually for both of them $z = 2$. In the 
other limit when the correlation length is of the order of $L$, then 
$\tau_{\rm max} \sim L ^4$. In this case $z = 4$.

When the system approaches criticality, one expects the correlation length 
$\xi$ to diverge. Using this fact, it is possible to estimate the critical 
temperature ($T_C$) which satisfies the following equation,
\begin{eqnarray}
\label{tc_eq} 
1-2d\gamma = 0.
\end{eqnarray}

For one dimensional system, $z = 2$ and accordingly $L = 2(z-1) = 2$. 
Here $\gamma = \frac 12 tanh~(2\beta J)$ (see Eq. (\ref{gamma_value})). 
In this case Eq. (\ref{tc_eq}) takes the following form:
$tanh~(2\beta J)=1$. This will be only satisfied when
$\beta \rightarrow \infty$. Therefore in this case $T_C =0$, in accordance
with the fact that the one dimensional Ising system behaves critically only
near to absolute zero temperature.

For two dimensional system (square lattice), $z =4$ and accordingly $L=6$. 
Here $\gamma = \frac{1}{32}[tanh~(6\beta J)+4 tanh~(4\beta J) 
+ 5 tanh ~(2\beta J)]$ (see Eq. (\ref{gamma_value})). 
In this case Eq. (\ref{tc_eq}) takes the following form:
$tanh~(6\beta J)+4 tanh~(4\beta J) 
+ 5 tanh ~(2\beta J) = 8$. Solution of this equation
gives $T_C = 2.493J/k_B$, whereas its exact value is know to be
$T_C = 2.269J/k_B$ \cite{onsager44}.

For three dimensional system (simple cubic lattice), $z = 6$ and accordingly 
$L=10$. Here $\gamma =  \frac{1}{512}[tanh~(10\beta J)+
8 tanh~(8\beta J) + 27 tanh ~(6\beta J) + 48 tanh~(4\beta J) +
42 tanh~(2\beta J)]$ (see Eq. (\ref{gamma_value})). In this
case Eq. (\ref{tc_eq}) takes the following form:
$[tanh~(10\beta J)+
8 tanh~(8\beta J) + 27 tanh ~(6\beta J) + 48 tanh~(4\beta J) +
42 tanh~(2\beta J)] = 256/3$. Solution of this equation
gives $T_C = 4.342J/k_B$, whereas its actual value is expected to be
about $T_C = 4.511J/k_B$ \cite{salman98,livet91,talapov96}.

We see here that the values of the critical temperatures ($T_C$) are in very good 
agreement with the actual ones, and an impressive improvement over the 
mean field values (where $T_C = zJ/k_B$ with $z$ = 2, 4 and 6 for one, two and 
three dimensional systems respectively). It is here encouraging to notice 
that our approach correctly captures the basic physics of the Ising model 
in different dimensions, viz., while the one dimensional system does not exhibit 
criticality at any finite temperature, the two and three dimensional systems do 
exhibit criticality at finite temperatures.

Before finishing this subsection, we briefly comment on scaling behavior of the
correlation length. Near to the criticality, the correlation length diverges as, 
$\xi \sim |\frac{T-T_C}{T_C}|^{-1/2}$ which can be seen by noting that 
$|1-2d\gamma| \sim |\frac{T-T_C}{T_C}|$. This shows that the critical exponent 
$\nu = 1/2$ (for $d$ = 2 and 3). 

\subsection{Steady state correlation function}

As we mentioned before, for a general case, physically appealing solution of 
Eq. (\ref{G_eq_d}) is difficult. It is though possible to quickly look into 
specific aspects of the equation, for example, by studying its steady state 
solution. If $P(r)$ is the function $G(r,t)$ for $t \rightarrow \infty$, 
then $P(r)$ should satisfy the following equation:
\begin{eqnarray}
\label{crltn_eq} 
\nabla^2 P(r) = \xi^{-2} P(r), 
\end{eqnarray}
where again the correlation length $\xi = \sqrt{\frac{\kappa}{D}} = 
\sqrt{\frac{\gamma}{1 - 2d \gamma}}$. The solution of this equation 
can be found in Ref. \cite{sahoo14}. A trial solution of the form 
$\frac{e^{-r/\xi}}{r^l}$ can be taken to find the desired solution for 
$P(r)$. Here $l$ is a constant to be determined; we find that $l$ = 0 and 1 
for $d$ = 1 and 3 respectively. For $d = 2$, the above trial form does not 
yield any solution of Eq. (\ref{crltn_eq}). For this special case, we take the
following trial form: $P(r) = S(r)e^{-r/\xi}$. It is easy to see that $S(r)$ 
satisfies the following equation:
\begin{eqnarray}
\label{twosol}
\xi r \frac{d^2 S(r)}{dr^2} + (\xi -2r) \frac{dS(r)}{dr} - S(r) = 0.
\end{eqnarray}
Solution of this equation can most easily be found by a trial series of the 
form $S(r) = \sum_{n=0}^\infty a_n r^n$. We obtain the following solution 
for the function:  
$S(r) = a_0 [1+ \sum_{n=1}^\infty \frac{(2n-1)!!}{(n!)^2}(r/\xi)^n]$. With 
this result, we now write down the solution for $P(r)$: 
\begin{eqnarray}
\label{P_sln}
P(r) = a_0e^{-r/\xi}\left\{ \begin{array}{l l}
1,& d = 1 \\
1+ \sum_{n=1}^\infty \frac{(2n-1)!!}{(n!)^2}(r/\xi)^n, & d = 2 \\
\frac{1}{r},&  d = 3.  
\end{array} \right.
\end{eqnarray}
The value of $a_0$ can be determined by a normalization condition (the value 
will be, of course, different for different dimensions). Near to the 
criticality ($\xi \rightarrow \infty$), it is easy to see that 
$P(r) \sim \mathcal{O}(1)$, $\mathcal{O}(1)$ and $1/r$ 
for one, two and three 
dimension respectively (we assume here $r \ll \xi$). 
This suggests that the values of the critical exponent $\eta$, defined  
as $P(r) \sim r^{-(d-2+\eta)}{\rm e}^{-r/\xi}$, are 1,
0 and 0 respectively for $d$ = 1, 2 and 3.

\subsection{Some additional remarks}
We would like to make some comments here before finishing this section. 
The diffusion constant $D$ (coefficient of the second order term in 
Eq. (\ref{Q_eq_d})) vanishes at criticality ($T$ = $T_C$). This shows that the 
diffusion process goes through a critical slow down near $T_C$. 
%\textcolor{red}{same paragraph to be continued?}
We see that at the criticality, Eq. (\ref{Q_eq_d}) reduces to a bidiffusion 
equation, $\frac{\partial}{\partial t} Q(\vec{r},t)= - \kappa \nabla^4 Q(\vec{r},t)$. 
It is easy to solve this equation, using Fourier transforms, and check that the 
average domain size grows as $L(t) \sim t^{1/4}$ \cite{krapivsky10}. By now it is 
though well established that $L(t) \sim t^{1/3}$ after a deep quench 
\cite{krapivsky10,bray94}. This contradictory results may not be surprising as 
at the criticality, due to critical slow down, we expect slower growth rate of 
domains.

\section{Conclusion}
The Kawasaki model is not exactly solvable (in any dimension) as the exchange rate 
($w_{jj'}$) involved in the calculations is highly nonlinear. To make the 
calculations tractable, in this paper we discussed a mathematical way to linearize 
$w_{jj'}$ in such a way that the mean squared error in satisfying the detailed 
balance condition is least. In the continuum limit, our approach leads to a 
Cahn-Hilliard equation of conservative dynamics. This establishes a connection
between the phenomenological Cahn-Hilliard theory and the Kawasaki's 
microscopic model for conservative dynamics. Advantage of our work is that it will 
help us anticipate how the conservative dynamics of an arbitrary Ising system depends 
on the temperature and the coupling constants. In particular, the critical 
temperatures estimated from the divergence of correlation length or the critical slow 
down are in very good agreement with the actual values.

\ack
SS acknowledges financial support by Prof. S. Ramasesha through his project from 
DST, India and SC acknowledges financial support from DST SwarnaJayanti project of 
Dr. Bedangadas Mohanty.


\section*{References}
\begin{thebibliography}{10}
\bibitem{glauber63}Glauber R J 1963 {\it J. Math. Phys.} {\bf 4} 294
\bibitem{kawasaki66}Kawasaki K 1966 {\it Phys. Rev.} {\bf 145} 224
\bibitem{sahoo14} Sahoo S and Ganguly S K 2014 {\it preprint} arXiv:1401.5412
\bibitem{scheucher88} Scheucher M and Spohn H 1988 {\it J. Stat. Phys.}
{\bf 53} 279
\bibitem{oliveira03} de Oliveira M J 2003 {\it Phys. Rev. E} {\bf 67} 066101
\bibitem{hase06}Hase M O, Salinas S R, Tom\'{e} T and de Oliveira M J 2006
{\it Phys. Rev. E} {\bf 73} 056117
\bibitem{krapivsky10}Krapivsky P L, Redner S and Ben-Naim E 2010
{\it A kinetic view of statistical physics} (Cambridge:
Cambridge University Press) 
\bibitem{bray94} Bray A J 1994 {\it Adv. Phys.} {\bf 43} 357
\bibitem{suzuki68} Suzuki M and Kubo R 1968 {\it J. Phys. Soc. Jpn.} {\bf 24} 51
\bibitem{puri09}Puri S 2009 {\it Kinetics of phase transitions}, ed S Puri  
and V Wadhawan (Boca Raton: CRC Press)
\bibitem{cahn59} Cahn J W and Hilliard J E 1959 {\it J. Chem. Phys.} {\bf 31} 688
\bibitem{cahn65} Cahn J W 1965 {\it J. Chem. Phys.} {\bf 42} 93
\bibitem{onsager44}Onsager L 1944 {\it Phys. Rev.} {\bf 65} 117
\bibitem{salman98} Salman Z and Adler J 1998 {\it Int. J. Mod. Phys. C}
{\bf 09} 195
\bibitem{livet91}Livet F 1991 {\it Europhys. Lett.} {\bf 16} 139
\bibitem{talapov96}Talapov A L and Bl\"{o}te H W J 1996
{\it J. Phys. A: Math. Gen.} {\bf 29} 5727

\end{thebibliography}
\end{document}